\documentstyle[12pt, aasms4]{article}
\begin{document}

\title{CO(4--3) and dust emission in two powerful high-z radio galaxies,
and  CO lines at high redshifts}

\author{Padeli P. Papadopoulos\altaffilmark{1}}

\author{H. J. A. R\"ottgering\altaffilmark{1}}

\author{P. P. van der Werf\altaffilmark{1}} 

\author{S. Guilloteau\altaffilmark{2}}

\author{Omont A\altaffilmark{3}}

\author{W. J. M. van Breugel\altaffilmark{5}}

\and

\author{R. P. J. Tilanus\altaffilmark{4}}

\altaffiltext{1}{Sterrewacht Leiden,  P. O. Box 9513,  2300 RA Leiden,
The Netherlands} 

\altaffiltext{2}{IRAM,  300 rue  de la  Piscine, F-38406  Saint Martin
d'H\'eres Cedex, France}

 \altaffiltext{3}{Institut d'  Astrophysique de Paris,  CNRS, 98bis Bd
Arago, F-75014 Paris, France}

\altaffiltext{4}{Joint  Astronomy Center,  Komohana  Street, Hilo,  HI
96720}

\altaffiltext{5}{Institute   of  Geophysics  and   Planetary  Physics,
Lawrence Livermore Laboratory, PO Box \hspace*{0.5cm} 808, Livermore, CA 94550}

\begin{abstract}

We report  the detection of sub-mm  emission from dust at  850 $\mu $m
and of  the $ ^{12}$CO J=4--3  line in the two  distant powerful radio
galaxies 4C 60.07  (z=3.79) and 6C 1909+722 (z=3.53).   In the case of
4C 60.07 the dust emission is also detected at 1.25 mm.  The estimated
molecular gas masses  are large, of the order  of $\sim (0.5-1) \times
10^{11}\rm  \ M_{\odot }$.   The large  FIR luminosities  ($\rm L_{\rm
FIR}\sim  10^{13}\ L_{\odot  }$) suggest  that we  are  witnessing two
major starburst  phenomena, while  the observed large  velocity widths
($\Delta \rm V_{\rm FWHM}\ga 500$  km s$ ^{-1}$) are characteristic of
mergers.  In the  case of 4C 60.07 the CO  emission extends over $\sim
30$ kpc  and spans a  velocity range of  $\ga 1000$ km s$  ^{-1}$.  It
consists of two distinct features with  FWHM of $\ga 550$ km s$ ^{-1}$
and  $\sim  150$~km~s$  ^{-1}$  and  line centers  separated  by  $\ga
700$~km~s$ ^{-1}$.  The least  massive of these components is probably
very gas-rich with potentially $\ga 60\%$ of its dynamical mass in the
form  of  molecular  gas.   The  extraordinary morphology  of  the  CO
emission  in this  object suggests  that it  is not  just  a scaled-up
version of  a local Ultra  Luminous Infrared Galaxy,  and it may  be a
formative stage of the elliptical host of the residing radio-loud AGN.
Finally  we briefly  explore  the effects  of  the wide  range of  gas
excitation  conditions  expected  for  starburst environments  on  the
luminosity of high-J CO lines. We conclude that in unlensed objects, CO
(J+1$\rightarrow $  J), J+1$>$3 lines  can be significantly  weak with
respect to CO  J=1--0 and this can hinder their  detection even in the
presence of substantial molecular gas masses.

\end{abstract}


\section{Introduction}

The successful detection of CO emission in many local ($\rm z\la 0.3$)
IRAS  galaxies revealed large  reservoirs of  molecular gas  mass that
fuel nuclear starbursts  and Active Nuclei (e.g.  Tinney  et al. 1988;
Sanders,  Scoville  \&  Soifer  1991;  Solomon  1997).   In  the  most
IR-luminous  galaxies  ($\rm  L_{\rm  FIR}\ga  10^{11}\  M_{\odot  }$)
gas-rich mergers and  interactions are thought to play  a crucial role
in causing the  rapid accumulation of molecular gas  in the center and
thus initiating the onset  of spectacular starbursts.  Subsequent high
resolution imaging  of the CO  emission in Ultra Luminous  IR Galaxies
(hereafter  ULIRGs) (e.g.   Scoville, Yun  \& Bryant  1997;  Downes \&
Solomon 1998) confirmed this  picture by demonstrating the presence of
molecular gas  with high mass surface densities  ($\rm \Sigma (H_2)\ga
5\times 10^3\ M_{\odot }\  pc^{-2}$) and large velocity widths ($\sim
500$  km  s$ ^{-1}$)  confined  within a  few  hundred  parsec in  the
nuclear~regions.

Besides being responsible for  some of the most spectacular starbursts
in  the local  Universe  the merging  process  is thought  to play  an
important role  in galaxy formation  at high redshift,  especially for
spheroidal systems.  The favorite formation model for these systems is
that  they   grow  from   the  hierarchical  clustering   of  gas-rich
``fragments''  where the  oldest  stars ($\gg  10$  Gyr) have  already
formed  (e.g.   Baron \&  White  1987;  White  1996).  The  subsequent
intense star formation rapidly consumes the gas mass and finally gives
rise  to a present-day  giant-elliptical galaxy  with its  red colors,
evolved stellar  population and large mass, but  devoid of substantial
amounts of  gas and dust.   Powerful high-z radio  galaxies (hereafter
HzRGs) may  be the progenitors of the  massive present-day ellipticals
hosting a radio-loud AGN.  Since  such galaxies seemed to have already
settled into  ellipticals (Best,  Longair, \& R\"ottgering  1997, 1998
and 1999) by $\rm z\sim 1$, it is natural to assume that HzRGs at $\rm
z>2$ with their frequently irregular morphologies (e.g.  Pentericci et
al.   1999) is where  merger-induced large  scale starbursts  may have
occured.

The   first  detection   of   CO  in   IRAS   10214+4724  at   z=2.286
 (Brown~\&~Vanden~Bout~1991, 1992) and  the subsequent detection of its
 mm/sub-mm continuum from dust (see Downes et al.  1992 and references
 therein)  initiated  ongoing  efforts  to  detect  CO  and  mm/sub-mm
 emission from  the copious  amounts of gas  and dust expected  in the
 high-z  counterparts  of  ULIRGs  and in  galaxies  undergoing  their
 formative starbursts.  The  large negative K-corrections expected for
 the thermal dust spectrum and the high-J CO lines (e.g.  Hughes 1996;
 van  der  Werf  \&  Israel  1996)  of  high-z  galaxies  as  well  as
 gravitational  lensing help to  make such  objects detectable  to the
 current  mm/sub-mm  instruments.    Lensing  amplifies  the  expected
 emission  but usually complicates  the gas  and dust  mass estimates.
 The galaxy IRAS 10214+4724 and many high-z CO/sub-mm luminous objects
 like the  QSO H1413+117 (Barvainis  et al.  1994), the  recent sub-mm
 selected galaxies (Frayer et al 1998; Frayer et al. 1999) and the BAL
 quasar APM 08279+5255 (Downes et al. 1999) are lensed.

Recently there have been detections  of large amounts of dust in HzRGs
(e.g. Hughes  1996; R\"ottgering et  al. 1998; Hughes \&  Dunlop 1998;
Best et  al. 1998), and a  systematic survey is now  been conducted to
detect  sub-mm emission  from dust  for all  $\rm z>3$  radio galaxies
(R\"ottgering et al.  1999) in  order to understand the large range of
FIR luminosities of these objects.   The most notable examples are the
extremely  FIR-luminous  radio galaxies  8C  1435+635  (Ivison et  al.
1998) and 4C~41.17 (Dunlop et al.  1994; Chini \& Kr\"ugel 1994) whose
properties do  not appear to  be influenced by  gravitational lensing.
In both cases the inferred gas masses are large enough ($\sim 10^{11}\
\rm M_{\odot }$)  to suggest that the large  FIR luminosities of these
objects  ($\sim 10^{13}\  \rm L_{\odot  }$) are  due to  the formative
starbursts.  However, despite systematic  efforts (Evans et al.  1996;
van Ojik 1997) no CO emission has been detected from these two objects
or any other  powerful HzRGs.  In this paper  we present the detection
of dust continuum  and CO J=4--3 line emission  in two powerful high-z
radio galaxies and discuss the  latter in the context of the earlier
unsuccessful  attempts  to  detect   CO  in  this  class  of  objects.
Throughout this  work we  assume $\rm H_{\circ  } = 75\  km\ sec^{-1}\
Mpc^{-1}$ and~$\rm q_{\circ } = 0.5$.

\section{Observations and data reduction}

\subsection{The SCUBA observations}
 
We used  the Sub-mm  Common User Bolometer  Array (SCUBA) at  the 15-m
James Clerk Maxwell Telescope  (JCMT)\footnote{The JCMT is operated by
the Joint  Astronomy Center  in Hilo, Hawaii  on behalf of  the parent
organizations  PPARC  in the  United  Kingdom,  the National  Research
Council of Canada and  the The Netherlands Organization for Scientific
Research.} in the photometry mode to observe 4C 60.07 on 1998 April 15
and 6C 1909+722 on 1997 October 4  and 5 as part of an ongoing program
to observe all  HzRGs at $\rm z>3$ (R\"ottgering  et al. 1999).  SCUBA
is  a dual  camera system  cooled to  $\sim 0.1$~K  allowing sensitive
observations  with two  arrays  simultaneously.  The  short-wavelength
array at 450 $\mu $m  contains 91 pixels and the long-wavelength array
at 850 $\mu $m 37  pixels.  The resolution is diffraction-limited with
$\rm HPBW\sim  15''$ (850 $\mu $m),  and $\rm HPBW\sim  8''$ (450 $\mu
$m).   For  a  full  description  of the  instrument  see  Holland  et
al. (1998).

We employed the  recommended rapid beam switching at  a frequency of 8
Hz and a  beam throw of $60''$ in azimuth.  The  pointing and focus of
the  telescope  were  monitored  frequently  using  CRL  618  and  QSO
0836+710, and the typical rms pointing error was $\sim 3''$.  Frequent
skydips  were  used to  deduce  the  atmospheric extinction.   Typical
opacities at  850 $\mu $m were $\tau  \sim 0.13$ for our  1998 run and
$\tau \sim  0.20$ for our  1997 run.  Beam  maps of Mars, CRL  618 and
Uranus were  used to derive  the gain $\rm C_{850}\sim  280\ mJy/beam\
mV^{-1}$.  The data were flatfielded, corrected for extinction and sky
noise  was  removed  before  they  were  co-added  following  standard
procedures outlined in  Stevens et al.  (1997).  The  estimated rms of
our measurements  is consistent with  what is expected from  the total
integration times  and the  NEFDs at the  sky conditions of  our runs.
The systematic uncertainty in the flux density scale at 850 $\mu $m is
of the order of $\sim 10\%$ (Holland, private communication).

\subsection{The IRAM interferometric observations}

We used  the IRAM Plateau de  Bure Interferometer (PdBI)\footnote{IRAM
is supported  by INSU/CNRS (France),  MPG (Germany) and  IGN (Spain).}
between  1998 April  20  and 1998  May~15  in the  D configuration  to
observe CO J=4--3 ($\nu _{\rm  rest} = 461.040$ GHz) in 6C~1909+722 at
z=3.534  (van Breugel  private communication)  in four  tracks ranging
from 4-14  hours each.  The same  configuration was also  used on 1998
November 8  and 29 to obtain two  8-hr tracks for 4C  60.07 at z=3.788
(Chambers  et al.   1996).  The  correlator  setup used  for the  3~mm
receivers involved 4  x 160 MHz modules covering  a total bandwidth of
560~MHz tunned in  single sideband. The band center  was positioned at
96.290 GHz  for 4C~60.07 and  101.770 GHz for  6C 1909+722.  The  1 mm
receivers were  used in  double sideband mode  with the  two remaining
correlator modules as backends to simultaneously observe the continuum
at  240~GHz.   Typical  SSB  system  temperatures were  $\sim  $120  K
(96.290~GHz) and $\sim $150~K (101.770  GHz), and at 240 GHz they were
$400-700$~K (DSB).   Bandpass calibration was obtained  using 3C 454.3
and   amplitude   and    phase   calibration   were   obtained   using
IAP$_{-}$0444+634 and NRAO  150.  For the 3 mm  receivers the residual
phase noise  was $\la  20^{\circ }$ and  for the amplitude  $\la 5\%$.
For the  1 mm receivers these  figures are $\la 30^{\circ  }$ and $\la
20\%$.  The flux density scale is accurate to within~$\sim 10\%$.

After  calibration  and  editing  the  data were  processed  with  the
standard NRAO AIPS software package. At 3 mm no continuum emission was
detected in  the line-free channels  and no continuum  subtraction was
performed.  The maps  were produced using the task MX  and in cases of
low  S/N,  CLEAN  was not  applied.   The  rms  noise  of the  maps  is
consistent  with the thermal  noise expected  for the  total observing
time and average $\rm T_{\rm sys}$ in the runs.

\section{Results}

Both radio galaxies are detected in  850 $\mu $m and CO J=4--3 and are
the  brightest sub-mm  objects found  in the  ongoing SCUBA  survey of
HzRGs at  $\rm z>3$ so far,  this being the main  reason for selecting
them for  the follow-up  sensitive CO observations.   In 4C  60.07 the
CO(4--3) emission is clearly resolved (Figure 1).

\placefigure{fig1}

A remarkable characteristic  of the CO emission in  this galaxy is its
large velocity range ($\ga 1000\ \rm km\ s^{-1}$) and its two distinct
components  with line  centers separated  by $\sim  700$~km~s$ ^{-1}$.
One component  has $\Delta \rm V_{\rm  FWHM} \ga 550\  \rm km\ s^{-1}$
(extending beyond the observed band), while the other is narrower with
$\Delta \rm V_{FWHM}\sim 150 \rm  \ km\ s^{-1}$.  The feature with the
narrow linewidth  coincides with the  position of the  suspected radio
core and thus the AGN, but  a significant part of the broad linewidth
component is  clearly offset  from it.  This  aspect of the  CO J=4--3
emission  is reminiscent  of the  CO J=5--4  emission detected  in the
radio-quiet quasar BR1202-0725 at z=4.69  (Omont et al.  1996) where a
broad and  a narrow  linewidth component are  also detected.   In that
object the narrow linewidth component  is centered on the AGN position
while the broad one is offset by $\sim 4''$.

  The velocity-integrated  map in Figure 2 shows  extended CO emission
with  two peaks  that are  $\sim  7''$ apart  ($\sim 30$  kpc at  $\rm
z=3.791$).  Towards  the same  region 1.25 ~mm  emission from  dust is
also detected, shown  in Figure 3.  Both the  CO and 1.25~mm continuum
maps are overlaid  on a map of the non-thermal radio  emission at 6 cm
(Carilli et al.~1997).

\placefigure{fig2}

\placefigure{fig3}

The radio  galaxy 4C 60.07 is  a powerful radio galaxy  ($\rm P_{4 \rm
cm}\sim 1.6\times 10^{27}$ W Hz$  ^{-1}$) with a Fanaroff-Riley II (FR
II) edge-brightened  double lobe morphology  (Fanaroff~\&~Riley 1974),
hence there  is the possibility  of a significant contribution  of the
non-thermal  emission  to  the   observed  1.25~mm  and  850  $\mu  $m
continuum.  Extrapolating  the non-thermal  flux density of  $\rm S_{6
\rm  cm}=19$  mJy  of  its  brightest component  with  its  associated
spectral   index    of   $\alpha    ^{3.6\rm   cm}   _{6\rm    cm}   =
-1.4$\footnote{spectral  index  defined  as  $\rm  S_{\nu}\propto  \nu
^{\alpha}$}  (De Breuck,  private communication)  yields  a negligible
contribution  ($\la 1\%$)  in  both 1.25  mm  and 850  $\mu $m  bands.
Further confirmation of the thermal origin of the 1.25~mm continuum is
offered by the  high resolution map in Figure 3  which shows that most
of this emission is not associated with the observed non-thermal radio
continuum and the  peak 1.25 mm brightness is  $\sim 4''$ (17.6 kpc)
offset from the  radio core where the AGN  probably resides (De Breuck
private communication).  Deep K$'$-band images obtained with Keck (van
Breugel et al.   1998) reveal faint emission towards  the weak eastern
radio component but  none from the regions with  the brightest 1.25 mm
continuum.   Since at  $\rm  z\sim 3.8$  the  K$'$-band is  rest-frame
B-band, this is consistent with the brightest 1.25 mm emission marking
the  place of  a  massive  gas/dust reservoir  with  large amounts  of
extinction.

In order to obtain the highest S/N maps for the two distinct CO J=4--3
features  we  averaged  all  the  appropriate  channels  and  the  two
resulting  maps  are shown  in  Figure~4  overlaid  with  the 1.25  mm
continuum.  From these maps it becomes obvious that the component with
the  largest dynamical and  molecular gas  mass is  closely associated
with the peak of the dust emission as expected.

\placefigure{fig4}

In the case of  the radio galaxy 6C 1909+722 a map  of the averaged CO
J=4--3 emission and  its spectrum are shown in  Figure~5. There it can
be seen that its linewidth  is also rather large. Extrapolation of the
non-thermal flux $\rm S_{\rm 6 cm}=59$ mJy with the observed power law
$\alpha   ^{6\rm  cm}  _{20\rm   cm}  =   -1.3$  (De   Breuck  private
communication) yields a non-thermal  contribution of $\sim 2\%$ at 850
$\mu $m.  All the observed  parameters for the two HzRGs are tabulated
in Table 1.

\placefigure{fig5}

\centerline{EDITOR: PLACE TABLE 1 HERE}

\section{Discussion}

The  two  galaxies  have  extended  radio emission  and  are  unlikely
candidates  for gravitationally  lensed objects.   In the  case  of 4C
60.07,  high-resolution radio images  show a  classical FR  II source.
Sensitive 6 cm (Carilli et  al.  1997), K-band (Chambers et al.  1996)
and deep K$ ^{'}$-band images (van  Breugel et al. 1998) do not reveal
any obviously  lensed features. For  6C 1909+722, HST images  at 7000$
\AA $  do not reveal  any lensed features either  (Pentericci, private
communication).   Henceforth  we  assume  that  the  two  sources  are
unlensed and organize the discussion as~follows:

1. Estimate  the molecular  gas  mass  implied  by the  CO  J=4--3
  luminosity assuming global gas  excitation conditions similar to the
  ones found in local starburst galaxies.

2. Briefly  explore the influence of  various galactic environments
   on  the  molecular gas  excitation  conditions  over large  scales.
   Special  emphasis   is  given  on   the  effects  of   the  various
   environments on the high-J CO lines and their detectability at high
   redshift.

3. Use  the  mm/sub-mm  data  to  find the  dust  masses  and  FIR
   luminosities and estimate the  $\rm M(H_2)/M_{\rm  dust}$  ratios,
   which we then compare to the values in the local Universe.

4. Discuss  the evolutionary  status of these  objects in  terms of
   their  gas mass  relative to  their  dynamic mass,  and their  star
   formation  rates and  efficiencies.   We focus  particularly on  4C
   60.07  where the  two  distinct molecular  gas components  strongly
   suggest a large scale  starburst event unlike the usually nuclear
   starbursts observed in local ULIRGs.

\subsection{Molecular gas content}

The  estimate of  molecular gas  mass  from the  CO J=1--0  luminosity
 involves the  so-called standard  conversion factor $\rm  X_{\rm CO}$
 (e.g.  Young  \& Scoville  1982; Bloemen 1985;  Dickman et  al. 1986;
 Young  \& Scoville  1991),  which is  ``calibrated'' using  molecular
 clouds in the Milky Way. From the CO(4--3) luminosity $\rm M(H_2)$ is
 then given by

\begin{equation}
\rm  M(H_2)  =  \left(\frac{\rm X_{\rm  CO}}{r_{43}}\right)  \frac{\rm
 c^2}{2\  k\ \nu_{43}  ^2}\left[  \frac{D^2 _L}{\  1+z\ }\right]  \int
 _{\Delta v} S_{\nu _{\rm obs}} d v,
\end{equation}

\noindent
where $\rm D_{\rm L}= 2\ c\ H_{\circ } ^{-1} (1+z -\sqrt{1+z})$ is the
luminosity  distance for  $\rm q_{\circ  } =0.5$,  $\nu  _{43}$~is the
rest-frame frequency of the CO  J=4--3 transition, $\rm r_{43}$ is the
(4--3)/(1--0)  line ratio  of the  area/velocity-integrated brightness
temperatures  and  $\rm  S_{\nu  _{\rm  obs}}$ is  the  observed  flux
density.  Substituting astrophysical units yields

\begin{equation}
\rm M(H_2) = 9.77 \times 10^9 \left(\frac{\rm X_{\rm  CO}}{r_{43}}\right)
\frac{\left(1+z-\sqrt{1+z}\right)^2}{1+z}\ \left[\frac{\int _{\Delta v}
 S_{\nu _{\rm obs}}dv}{\rm Jy\ km\ s^{-1}}\right]\ M_{\odot }.
\end{equation}

 The dependence of  $\rm X_{\rm CO}$ on the  ambient conditions of the
 $\rm H_2$ gas has been  extensively explored (e.g. Bryant \& Scoville
 1996;  Sakamoto 1996).   An important  recent result  is that  in the
 intense starburst environments of  ULIRGs the warm, diffuse gas phase
 dominating  the $ ^{12}$CO  emission does  not consist  of virialized
 clouds, hence  leading to  an overestimate of  $\rm M(H_2)$  when the
 standard  $\rm  X_{\rm  CO}\sim  5   \  M_{\odot  }\  (km  \  s^{-1}\
 pc^2)^{-1}$ is used (Solomon et al. 1997; Downes \& Solomon 1998).  A
 more  appropriate value  for such  ISM environments  and  hence their
 high-z  counterparts is $\rm  X_{\rm CO}\sim  1 \  M_{\odot }\  (km \
 s^{-1}\ pc^2)^{-1}$ (Downes \& Solomon~1998).

 A  plausible value  for $\rm  r_{43}$ can  be found  assuming  an ISM
 environment similar  to that of  a local ``average''  starburst.  The
 average  line   ratios  measured   towards  such  systems   are  $\rm
 r_{21}\approx 0.9$ (e.g.  Braine \& Combes 1992; Aalto et al.  1995),
 $\rm  r_{32}\approx  0.64$  (Devereux  et  al.  1994),  while  the  $
 ^{12}$CO/$  ^{13}$CO  J=1--0, 2--1  ratios  are  $\rm R_{10}  \approx
 R_{21}\approx 13$  (Aalto et al.   1995).  A Large  Velocity Gradient
 (LVG) code with  the aforementioned ratios as inputs  was employed to
 model  the physical conditions  of the  gas (e.g.   Richardson 1985).
 The conditions  corresponding to the  best fit are $\rm  T_{\rm kin}=
 50$~K,  $\rm n(H_2)\sim 10^3$  cm$ ^{-3}$  and $\rm  [CO/H_2]/dV/dr =
 3\times 10^{-5}\ (km  \ s^{-1} \ pc^{-1})^{-1}$, similar  to the ones
 deduced for  the high-z  starburst IRAS 10214+4724  (Solomon, Downes,
 Radford 1992).  For these conditions  it is $\rm  r_{43}=0.40$, which
 for  $\rm T_{\rm  CMB}=(1+z)\times 2.75\  K\sim 13\  K$,  is slightly
 enhanced to  $\rm r_{43}=0.45$.  This  ratio and $\rm X_{\rm  CO}=1 \
 M_{\odot  }\  (km  \  s^{-1}\ pc^2)^{-1}$  yield  $\rm  M(H_2)\approx
 8\times  10^{10}\ M_{\odot  }$ for  4C 60.07  and  $\rm M(H_2)\approx
 4.5\times  10^{10}\  M_{\odot }$  for  6C~1909+722.   These are  most
 likely lower limits  since even in ULIRGs the  ratio $\rm r_{43}$ can
 be significantly smaller but  not much larger.  Indeed the assumption
 of an ``average'' starburst excitation environment, while convenient,
 hides the fact that there is  a wide range of physical conditions for
 the  gas  in  IR-luminous  galaxies,   and  in  much  of  that  range
 CO(J+1$\rightarrow   $J),   $\rm  J+1>   3$   can  be   significantly
 weaker than CO(1--0).

As we shall see, the  potential faintness of the high-J CO transitions
 may partly  explain why attempts to  detect high-z CO  have been less
 successful than  the ones trying  to detect mm/sub-mm  continuum from
 dust.  Especially  for HzRGs two large systematic  searches (Evans et
 al.  1996;  van Ojik et al.   1997) gave null results.   On the other
 hand observations  of the  sub-mm continuum fare  better with  7 such
 objects detected at 850 $\mu $m (see van der Werf 1999 and references
 therein).  A~similar situation applies  also in other types of high-z
 objects.   While the  uncertainties of  the gas-to-dust  ratio  and a
 rising $\rm  L_{\rm FIR}/L_{\rm CO}$  ratio with FIR  luminosity (van
 der Werf 1999) can still  adequately account for this, the wide range
 of gas excitation observed in IR-luminous galaxies cannot but have an
 effect on the luminosity of CO lines especially at high J levels.

\subsection{The high-J CO transitions in ULIRGs at high z}

In the attempts to detect CO in high-z objects various transitions are
observed  depending  on  the  particular  redshift  and  the  receiver
used. For $\rm z\ga 2$ mostly CO(J+1$\rightarrow $J) with $\rm J+1\geq
3$ is  observed and  in the systematic  searches in  HzRGs transitions
with J+1=4-9  were routinely  observed (Evans et  al.  1996;  van Ojik
1997).  It  has been  argued that observing  higher J  transitions can
offset the dimming  due to the distance in high-z  objects in a manner
similar  to the negative  K-corrections of  the thermal  spectrum from
dust (van  der Werf  \& Israel  1996).  In this  picture the  warm and
dense gas  in an ``average'' starburst environment  thermalizes the CO
transitions  up to  J+1=6.  However  the luminosity  of the  high-J CO
transitions may be much smaller for two basic reasons, namely

1) The  presence of a  warm ($\rm  T_{\rm kin}=50-100$~K)  but diffuse
   ($\rm  n(H_2)\sim 10^2-10^3$ cm$  ^{-3}$) and  subthermally excited
   (for  $\rm J+1>  2$) gas  phase  that dominates  the $  ^{12}$CO
   emission in  ULIRGs (Aalto et  al.  1995; Downes \&  Solomon 1998).
   The most conspicuous such galaxy is Arp 220, frequently used as the
   standard ULIRG for comparison with high-z FIR-luminous sources.  In
   this source a low $\rm  r_{21}=0.53$ and high $\rm R_{10}>20$, $\rm
   R_{21}=18$ ratios are observed (Aalto et al.  1995; Papadopoulos \&
   Seaquist  1998), in  contrast to  the ``average''  starburst ratios
   mentioned~previously. Another important  characteristic of this gas
   phase is  the moderate  optical depths ($\tau  \sim 1-2$) of  the $
   ^{12}$CO (1--0) transition (Aalto et al. 1995).

2) A large reservoir of cold and/or subthermally excited gas extending
   beyond  the warm  starbursting nuclear  region of  an  ULIRG.  Such
   excitation gradients are observed  when large beams that sample the
   extended emission  are used  (Papadopoulos \& Seaquist  1998).  The
   associated cool dust  in the ULIRG VV 114  was imaged recently with
   SCUBA (Frayer  et al.   1999) and extremely  cold gas  ($\rm T_{\rm
   kin}=7-10$ K) is  inferred over large scales for  starbursts like IC
   5135 and NGC 7469 (Papadopoulos  \& Seaquist 1998). This gas phase,
   if  present, can  easily dominate  the {\it  global}  CO excitation
   especially in high-z systems where a beam of $\ga 5''$ at $\rm z\ga
   2$ corresponds to~$\rm L\ga 30$~kpc.

Here we must stress that the  effectiveness of a cold gas component in
suppressing the observed  global gas excitation is not  altered by the
higher  CMB temperature  at  high  z.  Indeed,  while  the higher  CMB
temperature enhances  the populations  of the high  J levels of  CO, it
also corresponds  to a higher background against  which the respective
lines must  be detected.  Moreover  the effective temperature  of cold
gas at  high z is not  simply the sum  of the temperature of  this gas
phase at z=0 and $\rm (1+z)\times 2.75$ K as one might naively assume.
This can  be demonstrated with  some simple arguments.   Obviously the
excitation temperature $\rm T_{\rm  exc}$ of any collisionally excited
line at any redshift is bounded as

\begin{equation}
\rm (1+z)\times T_{\rm cmb} \leq T_{\rm exc} \leq T_{\rm kin},
\end{equation}

\noindent
where $\rm  T_{\rm cmb}=2.75$ K  is the present epoch  CMB temperature
and $\rm T_{\rm  kin}$ is the gas kinetic  temperature.  If we further
assume  that on  large scales  in the  ISM of  a  galaxy thermodynamic
equilibrium exists  between gas and  dust, i.e.  $\rm T_{\rm  dust} =
T_{\rm kin}$,  then for  a dust emissivity  of $\alpha =2$  the energy
balance within a typical giant molecular cloud is described~by

\begin{equation}
\rm U_{\rm  ISRF} + U_{\rm  mech} = cooling\ rate  \propto \left[T_{\rm
dust}(z)^6- 2.75^6\times (1+z)^6 \right],
\end{equation}

\noindent
where $\rm  U_{\rm ISRF}$  is the energy  density of  the interstellar
radiation field  (O, B,  A stars) that  heats the grains  directly and
$\rm  U_{\rm  mech}$  is  the ``mechanical''  heating  energy  density
deposited to the molecular  clouds (e.g.  supernovae shocks, turbulent
cascade, cloud-cloud collisions) that first heats the gas and then the
dust.   These are  the  two  major heating  mechanisms  of an  average
molecular cloud and since, a)  more cooling processes are at play than
only  dust radiation  (e.g.  C$^+$,  CI,  CO lines),  b) usually  $\rm
T_{\rm   gas}\leq   T_{\rm  dust}$   (except   in   the  surfaces   of
UV-illuminated clouds, Hollenbach \& Tielens 1997),  it follows that
the dust  temperature yielded  by the last  equation will be  an upper
limit to the gas temperature.

 The aforementioned physical processes  responsible for the heating of
the ISM do  not depend on the particular value  of the CMB temperature
at  a given  z.   Thus the  last  equation, being  redshift-invariant,
allows us to  find the high-z equivalent temperature  of any given ISM
component from its temperature in the local Universe (see also Combes,
Maoli  \&  Omont 1999).   It  is  easy to  see  that  at  high z,  the
low-temperature  component  of  the  the  local ISM  has  $\rm  T_{\rm
dust}\rightarrow (1+z) \times T_{\rm  cmb}$ which also means (Equation
3) that lines become thermalized  at the expense of becoming invisible
against the enhanced CMB radiation field.

 Choosing a nominal  redshift of $\rm z=3.5$ as  representative of the
redshifts of  the two HzRGs we find  that for a warm  ISM component of
$\rm  T_{\rm  dust}(z=0)=50$  K   the  $\rm  T_{\rm  dust}(z=3.5)$  is
essentially identical.  However for a cold component of $\rm T_{\rm
dust}=10$~K  it is $\rm  T_{\rm dust}(z=3.5)  = 12.892$~K,  i.e.  just
$\sim  0.5$~K   above  the  $\rm  (1+z)\times  2.75\   K  =  12.375$~K
temperature of  the CMB at that  redshift.  In Table 2  we display the
expected line  ratios at  $\rm z= 3.5$  for three sets  of conditions,
namely  a)  an ``average''  starburst,  b)  diffuse  and warm  gas  of
moderate  $   ^{12}$CO  J=1--0  optical   depth,  and  c)   cold  gas,
representative  of the  phase  that  may exist  in  ULIRGs beyond  the
central starburst region.
\vspace*{0.5cm}

\centerline{EDITOR: PLACE TABLE 2 HERE}

This table shows that a  wide range is expected for CO(J+1$\rightarrow
 $J), J$>$2 line luminosities relative to CO J=1--0.  Several published
 $\rm  r_{32}=(3-2)/(1-0)$  line  ratios (Lisenfeld~et~al.~1996)  that
 sample the {\it global} CO emission  in ULIRGs, and a recent survey of
 this ratio in many nearby galaxies (Mauersberger et al.  1999) reveal
 a  large range  of values,  namely $\rm  r_{32} \sim  0.1-1$,  thus a
 similar  or  larger  range  is  to  be  expected  for  the  higher  J
 transitions.    Moreover   since   the   flux  density   ratio   $\rm
 S(J+1-J)/S(1-0)=(J+1) ^2\ r_{\rm J+1\ J}$ determines whether a high-J
 transition  is   easier  to   detect  than  J=1--0   (assuming  equal
 sensitivities),  it can  be  seen (Table  2)  that CO  J=3--2 is  the
 highest  transition for  which  this ratio  is  $\ga 1$  for all  the
 expected conditions.  Hence many non-detections of CO(J+1$\rightarrow
 $J), J$>$2  in high-z systems can  be due to  the globally sub-thermal
 excitation  of these  lines rather  than  a true  molecular gas  mass
 deficiency.  At  the same time this  makes the upper  limits for $\rm
 M(H_2)$ from such non-detections much less stringent.  For example if
 the CO J=5--4 line was  observed, and for $\rm X_{\rm CO}=1\ M_{\odot
 }\ (km \ s^{-1}\ pc^2)^{-1}$, the  upper limit in $\rm M(H_2)$ can be
 $\sim  1/0.025  \times  1/5=8$  times  larger that  what  is  usually
 reported for this line (e.g.  van Ojik~1998).

Finally for  gravitationally lensed objects with  underlying steep gas
excitation  gradients,  differential  amplification  can  render  line
ratios  useless  in deducing  the  average  gas  excitation unless  an
accurate  description of  the  lensing potential  is available.   Such
gradients are expected in nuclear  starbursts where gas warm and dense
enough to  excite high-J  CO lines is  more centrally confined  in the
immediate area of the starburst.  Recent  mapping of CO J=4--3 in M 51
and NGC 6946  (Nieten et al. 1999) showed that  even in more quiescent
galaxies the highly excited gas is strongly concentrated in the center.
Under  such circumstances differential  amplification can  enlarge the
effective size  of the high-J CO  emitting region with  respect to the
low-J one and thus alter the observed {\it global} line ratios towards
the ones expected  for warm and dense gas.   The frustrating aspect of
this effect is that it  can become progressively severe for lines that
are widely  separated in  J, i.e.  exactly  the ones whose  ratios are
most    sensitive   to   the    gas   excitation    conditions   under
normal~circumstances.

Clearly  more observations  of high-J  CO  lines in  local ULIRGs  are
needed in order to reveal the range of their {\it global} luminosities
and  relative  brightness  distributions   and  thus  allow  a  better
understanding of similar systems at high-z.

\subsection{Dust mass, the gas-to-dust ratio and $\rm L_{\rm FIR}$}

There are many uncertainties associated  with the estimate of the dust
 mass  from  mm/sub-mm  measurements  (e.g.  Gordon  1995),  with  the
 uncertainty of the  dust temperature being one of  the most important
 ones (Hughes  1996).  In the case  of 4C 60.07 the  detection of the
 dust continuum  both at 1.25  mm and 850  $\mu $m allows us  to place
 some  broad  constraints on  the  dust  temperature  as long  as  the
 emissivity  law is  $\alpha  >  1$.  Indeed,  for  an optically  thin
 isothermal reservoir of dust mass it is

\begin{equation}
\rm R(\alpha, T_{\rm d})=\frac{\rm  S_{850 \mu \rm m}}{\rm S_{1.25 \rm
mm}}=1.47^{\alpha        }\        \frac{\left(\rm        e^{81/T_{\rm
d}}-1\right)^{-1}-0.0022}{\left(\rm   e^{55/T_{\rm   d}}-1\right)^{-1}
-0.0162},
\end{equation}

\noindent
where $\rm T_{\rm d}$ is the dust temperature and $\alpha =1-2$ is the
exponent of the emissivity law.

For 4C 60.07 we find $\rm R(\alpha, T_{\rm d})=2.44\pm 0.73$. We adopt
an emissivity  law of $\alpha =2$,  which was deduced  in other high-z
objects, e.g.  IRAS  10214+4724 (Downes et al.  1992)  and 8C 1435+635
(Ivison et  al. 1998)  over a similar  rest-frame spectral  range. For
this  emissivity law the  $\pm 1\sigma  $ range  of the  observed $\rm
R(\alpha, T_{\rm d})$  yields a range of $\rm  T_{\rm d}\sim 20-50$ K.
Given the  starburst nature  of these two objects and their luminous
CO (4--3) line  (the J=4 level is $\sim 55$ K  above the ground state)
we assume a  dust temperature of $\rm T_{\rm d}=50$  K.  Then the dust
mass can be estimated from

\begin{equation}
\rm  M_{\rm dust}=\frac{\rm D^2  _L \  S_{\nu _{\rm  obs}}}{\left( 1+z
\right)\  k_{\rm   d}(\nu  _{em})}\  \left[B(\nu   _{\rm  em},  T_{\rm
d})-B(\nu _{\rm em}, T_{\rm cmb}(z))\right]^{-1},
\end{equation}

\noindent
where $\nu _{\rm  em}=(1+z) \nu _{\rm obs}$ is  the emitted frequency,
$\rm B(\nu,  T)$ is the Planck  function, $\rm T_{\rm  cmb}(z)$ is the
CMB  temperature at  redshift z,  and $\rm  k_{\rm d}(\nu  _{\rm em})=
0.04(\nu _{\rm em}/250 GHz)^2$ m$  ^2$ kgr$ ^{-1}$ is the adopted dust
emissivity (e.g.   Kr\"ugel, Steppe, \& Chini 1990).   For the assumed
$\rm T_{\rm d}$ the CMB term can be omitted with $\la 3\%$ error. Here
it  is  worth  noting  that  for  cold dust  this  term  can  be 
significant.  For example $\rm T_{\rm cmb}(\rm z=4)= 13.75$ K, hence a
dust component with  $\rm T_{\rm d}=15$ K {\it at  that redshift} is a
very cold component, just $\sim 1$ K above $\rm T_{\rm cmb}(\rm z=4)$.
An  observed wavelength of  1.25 mm  corresponds to  $\rm h  \nu _{\rm
em}/k\sim 57.4$ K at $\rm z=4$ and for this wavelength the CMB term is
$\sim 70\%$ of the dust term.

Assuming similar parameters  for the dust emission in  6C 1909+722 the
 850$\mu $m  flux density yields  $\rm M_{\rm dust} =  1.5\times 10^8\
 M_{\odot }$ for both objects.   This gives warm gas-to-dust ratios of
 $\rm   M(H_2)/M_{\rm   dust}  \approx   530$   (4C~60.07)  and   $\rm
 M(H_2)/M_{\rm dust} \approx 300$ (6C~1909+722), i.e. within the range
 of  the values  found for  local IRAS  galaxies (e.g.   Young  et al.
 1986; Stark et al.  1986; Young  et al.  1989) and ULIRGs (Sanders et
 al.  1991).  This  suggests that in a Universe at  $\sim 10\%$ of its
 current  age these  two  HzRGs already  have heavy-element  abundance
 comparable to the galaxies in  the contemporary Universe.  It is also
 worth  noting  that, while  many  uncertain  factors  enter into  the
 estimate of  the gas-to-dust ratio,  the values chosen for  them here
 are  the ones  considered  plausible from  the  study of  the ISM  in
 local~ULIRGs.

The FIR luminosities  of the two HzRGs were  estimated by assuming the
underlying spectrum of an optically thin, isothermal reservoir of dust
mass, hence

\begin{equation}
\rm L_{\rm  FIR}=\int_{0}^{\infty} L_{\nu  _{\rm em}} d\nu  _{\rm em}=
4\pi  M_{\rm dust}  \int_{0}^{\infty} k_{\rm  d}(\nu _{\rm  em}) B(\nu
_{\rm em}, T_{\rm d}) d \nu _{\rm em},
\end{equation}  

\noindent
 after using Equation 6 to  substitute $\rm M_{\rm dust}$ this finally
 yields

\begin{equation}
\rm  L_{\rm FIR}  = 4\pi\lambda(\alpha)  \ D^2  _{\rm  L}\ x^{-(\alpha
+4)}\left(e^{\rm x}-1\right) S_{\nu _{\rm obs }} \nu _{\rm obs},
\end{equation}

\noindent
where $\rm  x = h\nu _{\rm  em}/kT_{\rm d}$ and $\lambda  (\alpha )$ a
numerical constant that depends on the emissivity law. In astrophysical
units this becomes

\begin{equation}
\rm L_{\rm  FIR} = 2\times  10^7 \left(1+z-\sqrt{1+z}\right)^2 \lambda
(\alpha  )\ x^{-(\alpha  + 4)}\left(e^{\rm  x}-1\right) \left(\frac{\rm
S_{\nu  _{\rm  obs}}}{\rm  mJy}\right)\left(\frac{\nu _{\rm  obs}}{\rm
GHz}\right) \ L_{\odot }.
\end{equation}

For the assumed $\rm T_{\rm dust}=50$ K the last expression will yield
a lower limit  to the true $\rm L_{\rm FIR}$ since  in some ULIRGs the
optical  depth   may  become  significant  even   at  FIR  frequencies
(e.g. Solomon et al.  1997), and in a starburst environment dust can be
warmer  still. We  find $\rm  L_{\rm FIR}\approx  1.5  \times 10^{13}\
L_{\odot }$ for both HzRGs, a luminosity comparable to 8C~1435+635 and
4C 41.17  (Ivison et al.  1998  and references therein)  the other two
high-z radio galaxies detected in the sub-mm (rest-frame FIR) spectral
range whose properties do not appear to be influenced by gravitational
lensing.

\subsection{The evolutionary status of 4C 60.07 and 6C 1909+722}

The  usual  conclusion drawn  when  such  large  FIR luminosities  are
deduced for a  galaxy is that we are witnessing  it during a starburst
phase.  However many of the IR-luminous galaxies in the local Universe
and certainly the  two particular HzRGs also harbor  an AGN, thus part
of  the FIR luminosity  maybe due  to AGN-heated  dust.  Nevertheless,
recent studies of ULIRGs (Genzel  et al. 1998; Downes \& Solomon 1998)
reveal that most of them are powered by recently formed massive stars.
Also, even  when an AGN does  make a significant  contribution to $\rm
L_{\rm FIR}$, it is usually $\la  30\%$ (Genzel et al.  1998; Downes \&
Solomon 1998).  Therefore  the qualitative arguments behind converting
FIR luminosities to star formation rates are not likely to change even
if an AGN is present.  For  4C 60.07 the case for the starburst origin
of its  FIR luminosity is made stronger  by the fact that  most of its
1.25 mm dust emission is not emanating in the vicinity of the AGN.

The FIR  luminosity provides a  measure of the current  star formation
rate (SFR) according to

\begin{equation}
\rm SFR = 10^{-10}\ \Psi L_{\rm FIR} \ M_{\odot }\ yr^{-1},
\end{equation}

\noindent
where $\Psi \sim  1-6$, depending on the assumed  IMF (Telesco 1988 and
references therein). 

Adopting the most conservative value $\Psi \sim 1$, the FIR luminosity
of the  two HzRGs  implies $\rm SFR  \sim 1500\ M_{\odot  }\ yr^{-1}$.
Such  a  high  rate  of  star formation  can,  if  sustained,  produce
$10^{11}-10^{12}\rm \  M_{\odot}$ of stars in $\sim  0.06-0.6$ Gyr and
it  is comparable  to the  SFRs  found for  8C 1435+635  and 4C  41.17
(Ivison et al.  1998). The efficiency with which  such a burst converts
molecular gas into stars is  given by $\rm SFR/M(H_2)$ or equivalently
by $\rm  L_{\rm FIR}/M(H_2)$  (assuming the same  $\Psi $ for  all the
galaxies).   On many  occasions the  quantity $\rm  L_{\rm FIR}/L_{\rm
CO}$ is used instead, but since the $\rm X_{\rm CO}$ conversion factor
is $\sim 4-5$  times smaller in ULIRG systems the  latter ratio can be
misleading  when a  single  value for  $\rm  X_{\rm CO}$  is used  for
galaxies  spanning a  FIR range  from moderately  IR-luminous galaxies
($\rm  L_{\rm FIR}\sim 10^{10}\  L_{\odot }$)  to ULIRGs  ($\rm L_{\rm
FIR}\ga 10^{12}\ L_{\odot }$).

The star formation efficiencies  estimated are $\sim 190\rm \ L_{\odot
}\ M^{-1} _{\odot }$ (4C 60.07)  and $\sim 330\rm \ L_{\odot }\ M^{-1}
_{\odot  }$ (6C  1909+722), comparable  to the  ones found  for ULIRGs
(Solomon et al.  1997) once  $\rm M(H_2)$ has been estimated using the
same $\rm  X_{\rm CO}$ factor.   The implied star formation  rates and
efficiencies  of these galaxies  point towards  spectacular starbursts
occurring  at  high  redshift.    In  addition  the  large  linewidths
observed,  which in the  case of  4C~60.07 exceed  $\sim 1000$  km s$
^{-1}$, are  routinely observed towards  ULIRGs (e.g.  Solomon  et al.
1997) and    are    a    kinematic    signature   of    the    ongoing
mergers/interactions that trigger  their enormous starbursts (Sanders,
Scoville, \& Soifer~1991).

The tantalizing question raised when high-z starbursts are found is to
what extent the observed star formation episode is forming the bulk of
their eventual  stellar mass.  The  main difficulty in  answering this
important question  stems from the uncertainties  involved in deducing
the  total   gas  mass  from   CO  measurements.   If  we   assume  an
atomic-to-molecular gas  ratio of  $\rm M(HI)/M(H_2)\sim 2$  found for
IRAS galaxies (Andreani, Casoli, \&  Gerin 1995) and a large sample of
spirals (Casoli et  al.  1998), we obtain a total  gas content of $\rm
M_{\rm gas}=1.0\times 10^{11}(X_{\rm  CO}/r_{43})$ (4C 60.07) and $\rm
M_{\rm gas}=6.0\times 10^{10}(X_{\rm  CO}/r_{43})$ (6C~1909+722).  For
the  ``average''  starburst value  $\rm  r_{43}=0.45$  this gives  gas
masses of the order of $\sim (1.3-2.2)\times 10^{11}\ \rm M_{\odot }$.
These  are an order  of magnitude  higher than  typical gas  masses in
local  ULIRGs (e.g.  Downes  and Solomon  1998)  and constitute  $\sim
(15-20)\%$  of  the  total   stellar  mass  of  a  typical  elliptical
associated with the 3CR radio galaxies at $\rm z\sim 1$ (Best, Longair
\&  R\"ottgering  1998).   Thus   it  is  clear  that  these  enormous
starbursts still have a vast  reservoir of gas to be eventually turned
into stars.

The best  evidence yet that at  least in the  case of 4C 60.07  we are
witnessing an  extraordinary starburst and not a  scaled-up version of
an  ULIRG comes  from  the fact  that  in this  galaxy  the CO  J=4--3
emission from gas  is distributed over scales of  $\sim 30$ kpc.  This
is in contrast  with the local ULIRGs where most  of the molecular gas
and starburst activity  is confined within the central  $\la 1-5$ kpc.
Since starburst  activity is  usually co-extensive with  the molecular
gas  reservoir,  and this  will  be  particularly  true for  the  high
excitation CO  J=4--3 line,  it seems that  the starburst in  4C 60.07
occurs on galaxy-wide scales.

Equally intriguing  is the fact that  the CO emission  consists of two
distinct  components  widely separated  in  velocity  (Figures 1,  4).
Their   velocity-integrated  CO   flux  densities   of   $\ga  1.65\pm
0.35$~Jy~km~s$   ^{-1}$  (wide-linewidth  component)   and  $0.85\pm
0.20$~Jy~km~s$  ^{-1}$ (narrow-linewidth component)  yield  masses of
$\rm  M(H_2)\ga 5\times  10^{10}\ M_{\odot  }$ and  $\rm M(H_2)\approx
2.6\times   10^{10}\  M_{\odot}$   respectively   (Equation  2,   $\rm
r_{43}=0.45$).  We  derived size estimates for these  two components by
fitting both  the image  and the visibility  plane of the  CO emission
shown   in   Figure  4   with   an   underlying  gaussian   brightness
distribution. The  narrow-linewidth component appears  unresolved with
an  upper  limit  of  $\sim  4''$  while  the  wide-linewidth  one  is
marginally  resolved with  the largest  size being  $\sim 5''$.   In the
absence   of  adequate   spatial/kinematic   information  allowing   a
distinction  between  the  possible  geometrical arrangements  of  the
CO-emitting  gas we  assume that  the largest  estimated size  $\rm L$
corresponds to the diameter of a disk, hence its mass is given by

\begin{equation}
\rm M_{\rm  dyn }\approx \frac{\Delta V_{\rm FWHM} ^2\  \rm L}{\rm 2\
a_{\rm d}\  G\ sin^2\ i}  = 1.16\times 10^9  \left(\frac{\Delta V_{\rm
FWHM}}{100\rm   \    km\   s^{-1}}\right)^2   \left(\frac{\rm   L}{\rm
kpc}\right)\ \left(\rm sin^2\ i\right)^{-1}\ M_{\odot }
\end{equation}

\noindent
where  i is  the inclination  of the  disk and  $\rm a_{\rm  d}\sim 1$
 (Bryant \& Scoville 1996). 

For   the  wide-linewidth   component  the   largest   estimated  size
corresponds to $\rm L \sim 22$~kpc, which gives $\rm M_{\rm dyn }\sim
7.7\times 10^{11} (\rm sin^2\  i)^{-1}\ M_{\odot }$, comparable to the
mass of  a present-day  giant elliptical.  The  ratio of  the inferred
molecular to dynamic gas mass then is $\rm M(H_2)/M_{\rm dyn}\sim 0.06\
sin^2\  i$.   For  the   narrow-velocity  component  the  upper  limit
corresponds to  $\rm L\sim 17.5$ kpc  yielding a dynamic  mass of $\rm
M_{\rm dyn  }\la 4.6\times 10^{10} (\rm sin^2\  i)^{-1}\ M_{\odot }$.
This is  $\la 5\%$  of the mass  of a  typical giant elliptical  and a
comparison  with  its  molecular  gas mass  gives  $\rm  M(H_2)/M_{\rm
dyn}\ga  0.60\  sin^2\ i$.   Thus,  geometrical  factors aside,  this
component  is significantly  richer  in molecular  gas  than the  more
massive one and it maybe the  one where star formation has yet to form
the bulk of its eventual stellar mass.

It  is  easy  to   contrive  a  combination  of  different  excitation
properties,  $\rm X_{\rm CO}$  values, and  inclinations so  that $\rm
M(H_2)/M_{\rm dyn}$ in one or both molecular gas components is altered
significantly.  Still  the fact that  remains unaltered is that  in 4C
60.07, unlike  in a typical  ULIRG, the intense star  formation occurs
over  large  scales  in   two  spatially  and  kinematically  distinct
molecular gas  reservoirs.  This brings  in mind the scenario  for the
formation  of  a  giant  elliptical  at high  redshift  where  several
star-forming  clumps merge  to  eventually form  the  galaxy. In  this
picture a  gas-rich low mass  clump like the  one seen in 4C  60.07 is
still in the process of  merging and vigorous star formation while the
higher mass  object has already formed  most of its  stars.  Of course
the possibility  that the wide-linewidth component  itself consists of
several  virialized gas-rich clumps  that do not necessarily constitute
a bound system cannot be discarded in the light of the present data.

For 6C  1909+722 the CO emission  is unresolved with a  size $\la 4''$
($\rm  L\la 18$  kpc),  which  for the  observed  linewidth (Table  1)
corresponds  to $\rm  M_{\rm  dyn}\la 5.85\times  10^{11} (\rm  sin^2\
i)^{-1}\ M_{\odot  }$ and $\rm M(H_2)/M_{\rm dyn}\ga  0.08\ sin^2\ i$.
Thus, while there  is little doubt that this galaxy  is a starburst, it
can  still be  a high-z  counterpart  of an  ULIRG where  most of  its
eventual stellar  population has already  formed and the  intense star
formation is confined in its central region.

Finally, the case  of 4C 60.07 makes it clear  that when searching for
high-z  CO lines  a  large  velocity coverage  is  necessary not  only
because of their often uncertain redshift,  but in order to be able to
detect various components  that may be far apart  in velocity.  Higher
resolution  observations  are  necessary   in  order  to  reveal  more
structure  in the  molecular gas  reservoir of  this galaxy  and allow
better constraints  of its dynamical  mass.  However the  faintness of
the CO emission observed here may hinder such efforts and a systematic
study  of such  objects  may have  to  await the  advent  of the  next
generation  mm interferometers  with  significantly larger  collecting
areas.

\section{Conclusions}

In this  paper the detection of  mm/sub-mm emission from  dust and the
first detection of CO J=4--3 in two powerful high-z radio galaxies has
been presented.   The analysis of the  data leads us  to the following
conclusions:

1. Using the most conservative  CO/H$_2$ conversion factor (1/5 of the
galactic value) the CO J=4--3 emission implies molecular gas masses of
the order of $\rm M(H_2)\sim (0.5-1)\times 10^{11}\rm \ M_{\odot }$ in
the  two  high-z  powerful  radio galaxies.   Their  sub-mm  continuum
corresponds to large FIR  luminosities ($\sim 10^{13}\rm \ L_{\odot
}$) which imply that we  are witnessing intense starburst phenomena at
$\rm z\sim 3.5-3.8$ converting the aforementioned gas mass into stars.

2. The wide  range of the  molecular gas excitation expected  in Ultra
  Luminous IR  galaxies over large  scales is briefly explored  and we
  conclude that observing objects  with similar ISM conditions at high
  z  using  high-J CO  transitions  may  in  some cases  hinder  their
  detection.  This can partly  explain explain why, despite systematic
  efforts, CO detections of unlensed objects are fewer than detections
  of their mm/sub-mm  continuum from dust.  Therefore it  is still too
  early  to  draw  any  general  conclusions about  the  abundance  of
  molecular gas in high-z systems from the data currently available in
  the literature.

3. The  estimated  molecular gas-to-dust  ratios  found  in these  two
   objects  are  in the  range  found  for  the local  IRAS  galaxies,
   revealing that  in both of them most of the heavy  elements have
   already been produced at present-day abundances.

4. In 4C 60.07  the CO emission and presumably  the starburst activity
  implied by  its large FIR  luminosity is distributed over  scales of
  $\sim 30$  kpc and  consists of two  distinct components  spanning a
  total velocity  range of $\ga 1000$~km~s$ ^{-1}$.   Thus this galaxy
  does not seem to be just a scaled-up high-z version of a local Ultra
  Luminous  IR galaxy  where most  of the  starburst activity  and the
  accompanying bright CO/dust emission  are confined to a more compact
  region  in  the  center.   A  plausible  scenario  is  that  we  are
  witnessing the ongoing formation event of a giant elliptical galaxy,
  the future host of the residing radio-loud AGN.

\subsection{Acknowledgments}

We  acknowledge the  IRAM  staff from  the  Plateau de  Bure and  from
Grenoble for  carrying the observations  and help provided  during the
data reduction.   P.  P.  Papadopoulos would like  to thank especially
Dieter N\"urnberger and Anne Dutrey  for their patient help during the
data reduction in Grenoble and Jessica Arlett for providing us the LVG
code. We thank the referee David  Sanders as well as Simon Radford for
valuable comments, Chris Carilli for making available the 6 cm maps of
4C  60.07 and  Carlos De  Breuck for  help with  the radio  images and
stimulating  discussions.  The work  of  W.  v.  B. at  IGPP/LLNL  was
performed  under the  auspices of  the US  Department of  Energy under
contract W-7405-ENG-48.  P. P. P.   is supported by the ``Surveys with
the  Infrared  Space Observatory''  network  set  up  by the  European
Commission under contract ERB FMRX-CT96-0086 of its TMR programme.

{}

\clearpage

\figcaption{Naturally weighted channel maps of  CO J=4--3 for 4C 60.07
at  a   spatial  resolution  of  $8.9''\times   5.5''$  and  frequency
resolution of  35 MHz ($\sim  $ 110 km  s$^{-1}$). The contours  are $
(-3, -2, 2, 3, 4, 5, 6)\times  \sigma $, with $\sigma = 0.55\ \rm mJy\
beam^{-1}$. The FWHM of the restoring beam is shown at the bottom left
of the first channel.  The cross marks the position of the radio core,
 RA (J2000):  05 12 55.147, Dec (J2000): +60 30  51.0 (De Breuck
private communication)
\label{fig1}}

\figcaption{Uniform weighted, velocity-integrated  map of CO J=4--3 in
4C~60.07 (contours)  overlaid on  a 6 cm  map (greyscale) at  a common
resolution of $7.1''\times 5.2''$ The contours are $(2, 3, 4, 5)\times
\sigma $,  with $\sigma = 0.30$ Jy  beam$ ^{-1}$ km s$  ^{-1}$ and the
greyscale is: 1.5-15 mJy beam$ ^{-1}$.  The FWHM of the beam is shown
at  the bottom  left.  {\it  Note:} CLEAN  was not  applied in  the CO
J=4--3  uniform-weighted  channel  maps  because  of the  low  S/N  in
individual channels.  The cross marks  the position of the  radio core
(see Figure 1).
\label{fig2}}

\figcaption{Naturally weighted  map of the 1.25  mm continuum emission
of 4C~60.07 (contours) overlaid  to 6 cm continuum emission (greyscale)
at a common resolution of $3.7''\times 1.9''$.  The contours are $(-3,
-2, 2, 3, 4, 5)\times \sigma  $, with $\sigma = 0.6$~mJy~beam$ ^{-1}$ and
the  greyscale range is  0.6--12~mJy~beam$ ^{-1}$.   The FWHM  of the
restoring  beam is  shown  at the  bottom  left. The  cross marks  the
position of the radio core (see Figure 1).
\label{fig3}}

\figcaption{Naturally  weighted CO  J=4--3 maps  of the  two kinematic
components of  4C~60.07 (contours) overlaid  to its 1.25  mm continuum
(greyscale),  at  a  common  resolution  of  $8.9''\times  5.5''$.  The
velocity offset  is denoted at  the upper right,  and the FWHM  of the
line at the bottom  right. The contours are $ (-3, -2,  2, 3, 4, 5, 6,
7)\times  \sigma $, with  $\sigma =  0.3$ mJy  beam$ ^{-1}$  (top) and
$\sigma = 0.4$  mJy beam$ ^{-1}$ (bottom), and  the greyscale range is
1.3--5.85 mJy  beam$ ^{-1}$.  The FWHM  of the CLEAN beam  is shown at
the bottom left.  The cross marks the position of  the radio core (see
Figure 1).
\label{fig4}}

\figcaption{The CO  J=4--3 spectrum of 6C 1909+722  (top) with $\Delta
 \nu  _{\rm chan}=50$ MHz  \ ($\sim  145$~km~s$ ^{-1}$)  and naturally
 weighted  maps  of  the  CO   J=4--3  emission  centered  at  $\nu  =
 101.725$~GHz (middle) and $\nu = 101.975$~GHz (bottom).  The averaged
 frequency interval in both maps is $\Delta \nu = 200$ MHz ($\sim 590$
 km s$ ^{-1}$)  and the contours are $  (-3, -2, 2, 3, 4, 5,  6, 7, 8,
 9)\times \sigma $,  with $\sigma = 0.3$ mJy  beam$ ^{-1}$.  The cross
 marks the  map center (see  Table 1) which  is within 0.5$''$  of the
 position of  the radio core (De Breuck,  private communication).  The
 beam  size is  $7.65''\times 7.11''$  and its  FWHM is  shown  in the
 bottom map.
\label{fig5}}

\newpage

\centerline{\large Table 1}
\vspace*{1.0cm}
\centerline{\large Observational parameters}
\begin{center}
\begin{tabular}{ c c c }\hline\hline

Parameter & 4C 60.07 & 6C 1909+722 \\ \hline
RA  (J2000)$ ^{\rm a}$ & 05$ ^{\rm h}$ 12$ ^{\rm m}$ 54$ ^{\rm s}$.80 & 19$ ^{\rm h}$ 08$ ^{\rm m}$ 23$ ^{\rm s}$.70\\
Dec (J2000)$ ^{\rm a}$ & +60$ ^{\circ }$ 30$'$ 51.7$''$ & +72$ ^{\circ }$ 20$'$ 11.8$''$ \\
$\rm z_{\rm co}$ & $3.791$ & 3.532 \\
$\rm I_{\rm CO}$ (Jy km s$ ^{-1}$) & $2.50\pm 0.43$ & $1.62\pm 0.30$ \\
$\Delta \rm V _{\rm FWHM}$ (km s$ ^{-1}$) & $\ga 1000$ & $530\pm 70$ \\
$\rm S_{1.25\rm mm}$ (mJy) & $4.5\pm 1.2$ & $\leq 2$ (2$\sigma $) \\
$\rm S_{3\rm mm}$ (mJy)    & $\leq 0.5$ (2$\sigma $) & $\leq 0.6$ (2$\sigma $)\\
$\rm S_{850\mu \rm m}$ (mJy) & $11.0\pm 1.5$ & $13.5\pm 2.8$\\ \hline
\end{tabular}
\end{center}

\noindent
\hspace*{3cm} $ ^{\rm a}$ Coordinates of the image center.

\newpage

\centerline{\large Table 2}
\vspace*{1.0cm} 
\centerline{\large ISM environments and line ratios at $\rm z= 3.5$}
\begin{center}
\begin{tabular}{| c | c | c | c |}\hline\hline

Line ratios & ``Average'' starburst  & Diffuse warm gas & Cold gas \\ \hline\hline
 J=2--1     & 0.95--1.12                & 0.64--1.06        & 0.40--0.61   \\
 J=3--2     & 0.73--1.07                & 0.30--0.92        & 0.14--0.29 \\
 J=4--3     & 0.45--1.04                & 0.10--0.74        & 0.035--0.11 \\
 J=5--4     & 0.21--0.99                & 0.025--0.52       & 0.005-0.02   \\
 J=6--5     & 0.065--0.93               & 0.006--0.31  & $< 10^{-3}$--0.002  \\
 J=7--6     & 0.015--0.84               & 0.0015--0.15      & $< 10^{-3}$ \\
 J=8--7     & 0.002--0.69               & $<10^{-3}$--0.05 & $< 10^{-3}$  \\
 J=9--8     & $<10^{-3}$--0.50          & $<10^{-3}$--0.013 & $< 10^{-3}$ \\ \hline
\end{tabular}
\end{center}

Note.--- All the line ratios are normalized by the
 $ ^{12}$CO J=1--0 brightness, and the  
\hspace*{0.6cm} assumed CMB temperature is
 $(1+z)\times 2.75$ K = 12.375 K.

{\it ``Average'' starburst:} $\rm  T_{\rm  kin}=50-100$ K,
 $\rm n(H_2)=10^3-10^4$ cm$ ^{-3}$,\\
\hspace*{0.8cm} $\rm [CO/H_2]/(dV/dr)= 3\times 10^{-5}$ (km s$ ^{-1}$ pc$ ^{-1}$)$ ^{-1}$.
\vspace*{-0.3cm}                           

{\it Diffuse warm gas:} $\rm   T_{\rm  kin}=50-100$ K,
 $\rm n(H_2)=10^2-10^3$ cm$ ^{-3}$,\\
\hspace*{0.8cm} $\tau \sim 1-2$ for $ ^{12}$CO J=1--0.
\vspace*{-0.3cm} 

{\it Cold gas:} $\rm T_{\rm kin}\sim 13$ K (see text),
 $\rm n(H_2)=10^2-10^3$ cm$ ^{-3}$,\\
\hspace*{0.8cm} $\rm [CO/H_2]/(dV/dr) = 3\times 10^{-5}$ (km s$ ^{-1}$ pc$ ^{-1}$)$ ^{-1}$.

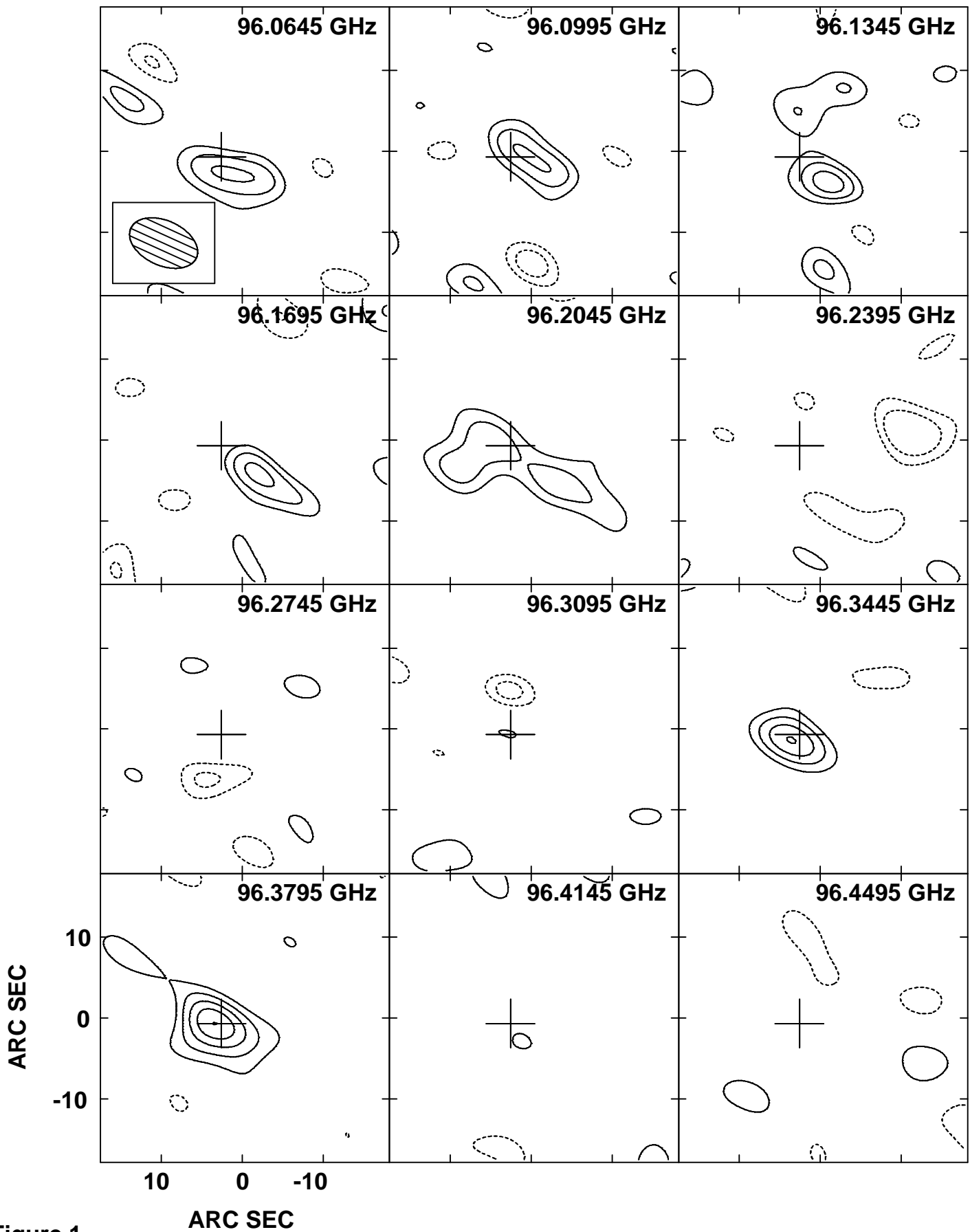
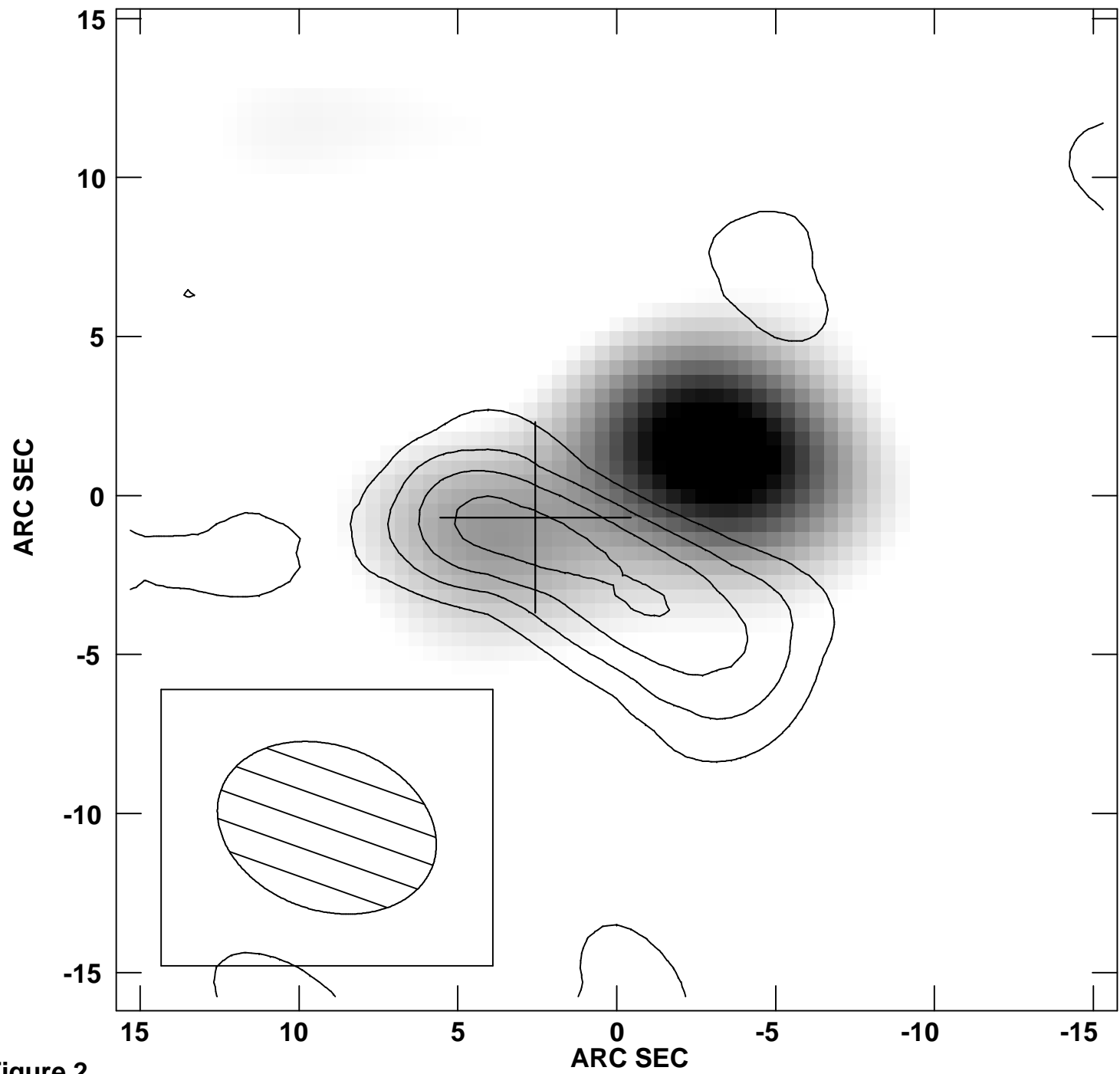
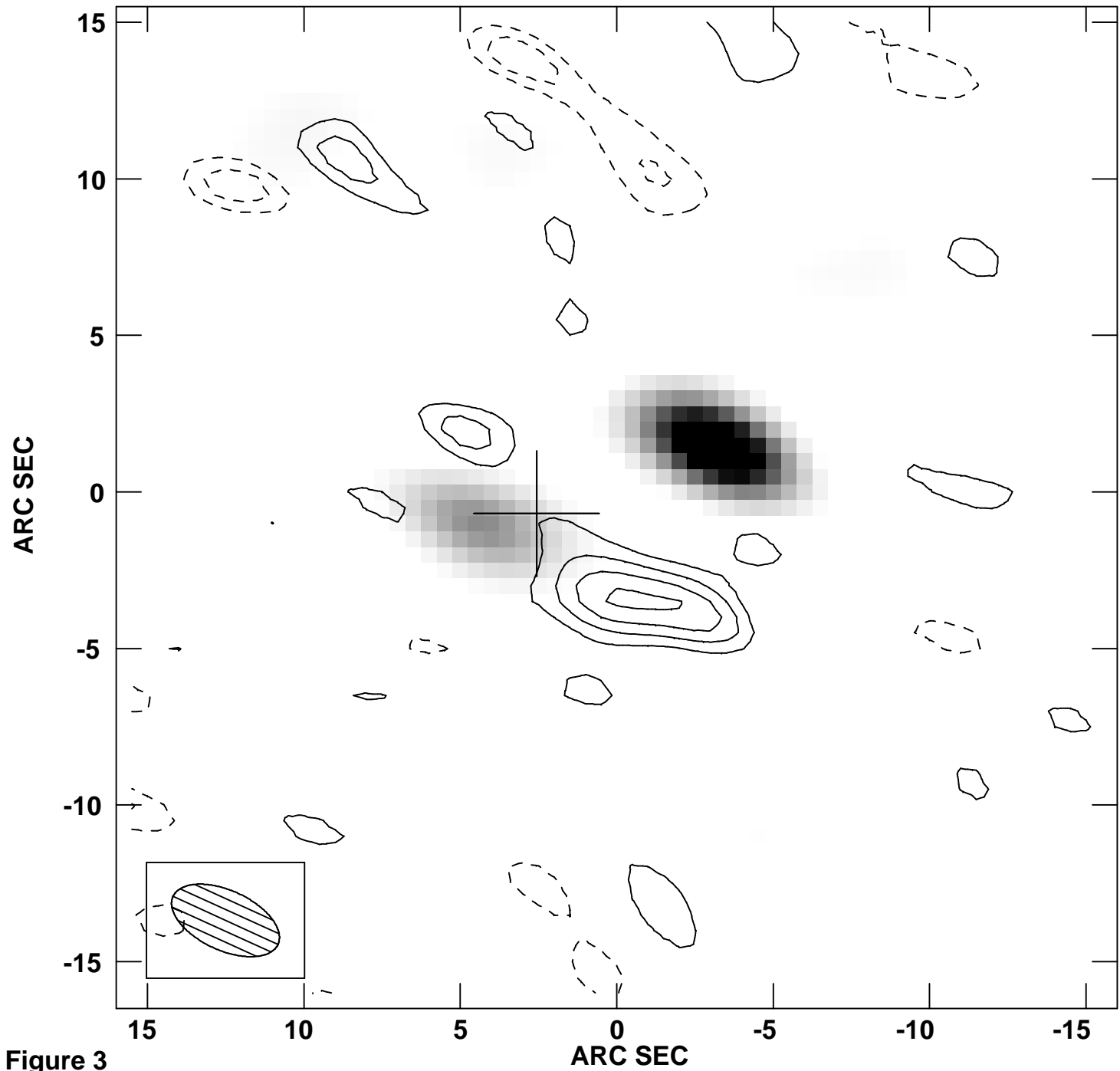
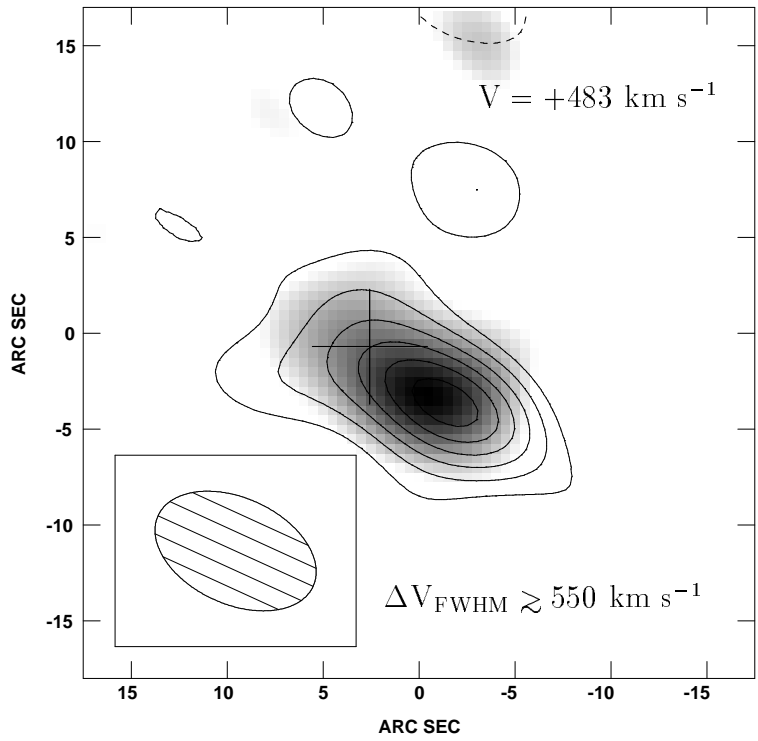
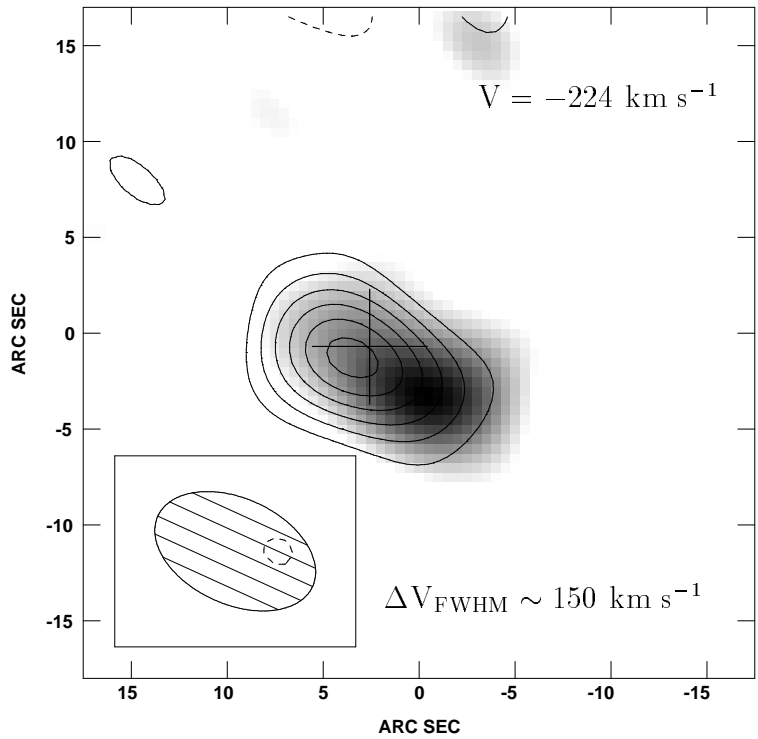
\end{document}